\documentclass[12pt]{iopart}

\ifx\pdfoutput\undefined
  \usepackage[dvips]{graphicx}
  \usepackage[dvips]{color}
\else
  \usepackage[pdftex]{graphicx}
  \usepackage[pdftex]{color}
\fi

\usepackage{iopams}
\usepackage{setstack}
\usepackage{latexsym}
\usepackage{units}
\usepackage{ulem}

\begin{document}

\title{Ultra-high-ohmic microstripline resistors for Coulomb blockade devices}

\author{Sergey V. Lotkhov}
\address{Physikalisch-Technische Bundesanstalt, Bundesallee
100, 38116 Braunschweig, Germany} \ead{Sergey.Lotkhov@ptb.de}

\date{\today}

\begin{abstract}
In this paper, we report on the fabrication and the low-temperature
characterization of extremely high-ohmic microstrip resistors made of a
thin film of weakly oxidized titanium. Nearly linear voltage-current
characteristics were measured at temperatures down to $T \sim
\unit[20]{mK}$ for films with sheet resistivity up to as high as $\sim
\unit[7]{k\Omega}$, i.e. about an order of magnitude higher than our
previous findings for weakly oxidized Cr. Our analysis indicates that such
an improvement can help to create an advantageous high-impedance
environment for different Coulomb blockade devices. Further properties of
the Ti film addressed in this work show a promise of low-noise behavior of
the resistors when applied in different realizations of the quantum
standard of current.

\end{abstract}

\submitto{\NT}

\maketitle

\section{Introduction}

Experimental observation and electronic applications of quantum tunneling
phenomena, in many cases, put forward the requirement of efficient
decoupling of the tunneling system from the environmental degrees of
freedom responsible for dissipation \cite{CaldeiraLeggett,IngoldNazarov}.
This implies, for the practical devices, that low noise operation should
be possible for the tunneling systems embedded into the DC biasing
circuitry with an effective output impedance $Re \left[Z(\omega)\right]
\gg R_{\rm Q} \equiv h/4e^2 \approx \unit[6.45]{k\Omega}$ far exceeding
the resistance quantum $R_{\rm Q}$ up to the frequency and energy ranges
relevant for the particular transport mechanisms
\cite{LikharevZorin,Pekola-NaturePhysics2008,MooijNazarov}.

For a mesoscopic tunneling system, one of the possibilities of
constructing a high-ohmic bias is to co-fabricate it on-chip with an
adjoining microstrip of a high-resistivity thin-film material. For
example, successful realizations of quantized charge transport have been
demonstrated in ultrasmall tunnel junctions, see, e.g., Refs.~
\cite{KuzminHaviland,LotkhovRpump,LotkhovBlochArray,Maibaum2012,CamarotaMetrologia},
and superconducting nanowires (as quantum phase slip junctions)
\cite{HongistoZorin} with the help of a few micrometer-long weakly
oxidized Cr resistors of a typical cross-section $S \sim \unit[10 \times
100]{nm^2}$ (thickness $\times$ linewidth) with the resistances per unit
length up to $r \sim \unit[5-10]{k\Omega/\mu m}$. A detailed analysis,
see, e.g., Ref.~\cite{ZorinJAP2000}, and the experimental data obtained so
far \cite{LotkhovBlochArray,Maibaum2012,HongistoZorin} indicate, however,
a significant shunting effect due to the stray capacitances of the
microstrip, typically $c \approx \unit[60]{aF/\mu m}$ \cite{ZorinJAP2000},
so that for the frequencies exceeding the roll-off value

\begin{equation}
\omega_0 = \frac{r}{2cR_0^2},\label{eq:rolloff}
\end{equation}
the on-chip resistor of a certain length $l$ and a designated impedance
$Z(0)= R_0 = rl$ must be considered as a distributed RC line rather than
as a simple lumped element (see, e.g., Ref.~\cite{IngoldNazarov}). In the
high-frequency limit, the impedance of the line decays approximately as a
square root of frequency

\begin{equation}
Re Z(\omega) \approx \sqrt{\frac{r}{2\omega c}},\label{eq:decay}
\end{equation}
descending below the resistance quantum $R_{\rm Q}$, i.e., beyond the
high-impedance case,  for the frequencies above $\omega_{\rm c} = \omega_0
\times (R_0/R_{\rm Q})^2$.

Given the due value of $R_0 \gg R_{\rm Q}$ and the value of $c$ fixed by
the microstripline design, equation~\ref{eq:rolloff} demonstrates an
important role of the unit length resistance $r$ for developing high-ohmic
environments effective in a wide frequency range. Unfortunately,
fabricating thinner or stronger oxidized Cr resistors often results in
irreproducible granular films exhibiting a pronounced Coulomb blockade of
a net current at low temperatures \cite{KrupeninChromium}. In the present
work, we report on resistivity behavior of films of titanium evaporated in
the presence of oxygen at low pressure. The choice of Ti is inspired by
its relatively high resistivity value, about $\unit[40]{\mu\Omega \times
cm}$ \cite{Webelements}, while its affinity to oxygen can be utilized for
avoiding the superconductivity of Ti below the transition temperature
$T_{\rm c} = \unit[0.4]{K}$, as well as for achieving even higher film
resistivity. At low temperature, we demonstrate a linear
current-to-voltage dependence for the Ti resistors with $r\le
\unit[70]{k\Omega/\mu m}$ being about an order of magnitude higher than
the corresponding values for the Cr films, which makes Titanium an
attractive material choice for resistively-biased Coulomb blockade devices
based on both tunnel- and quantum phase slip junctions.

\section{Experiment and Discussion}

\begin{figure}[]
\centering%
\includegraphics[width=0.8\columnwidth]{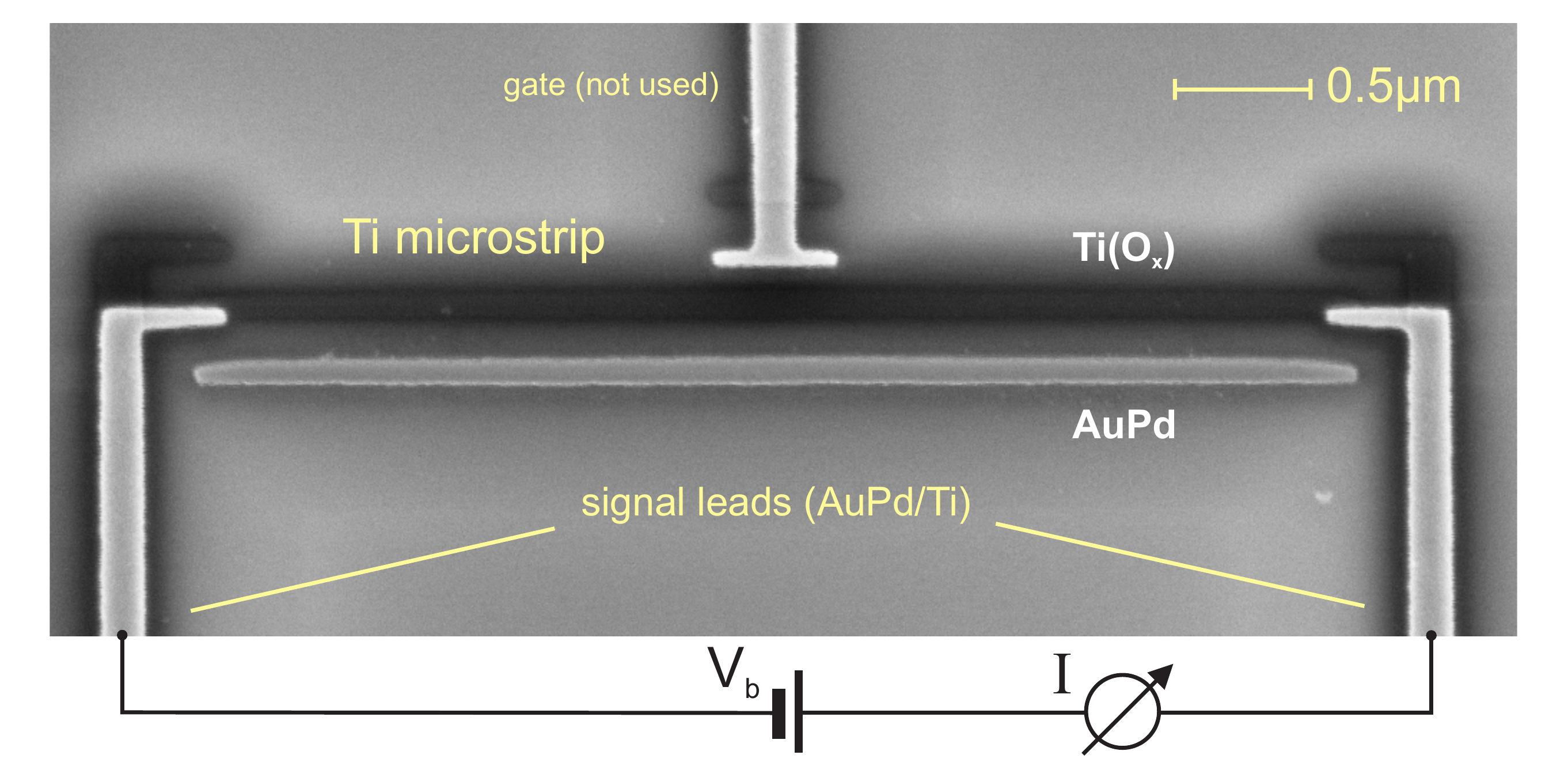}
\caption{SEM image of the $\unit[10]{nm}$ thick titanium resistor contacted by two $\unit[40]{nm}$ thick leads of AuPd.
The structure consists of two metallic replicas (Ti and AuPd) evaporated onto a tilted substrate at two different angles through the same hanging mask.
The gate electrode was not used in the present experiment. See the text for more details.}
\label{SEM}
\end{figure}

\begin{figure}[]
\centering%
\includegraphics[width=0.8\columnwidth]{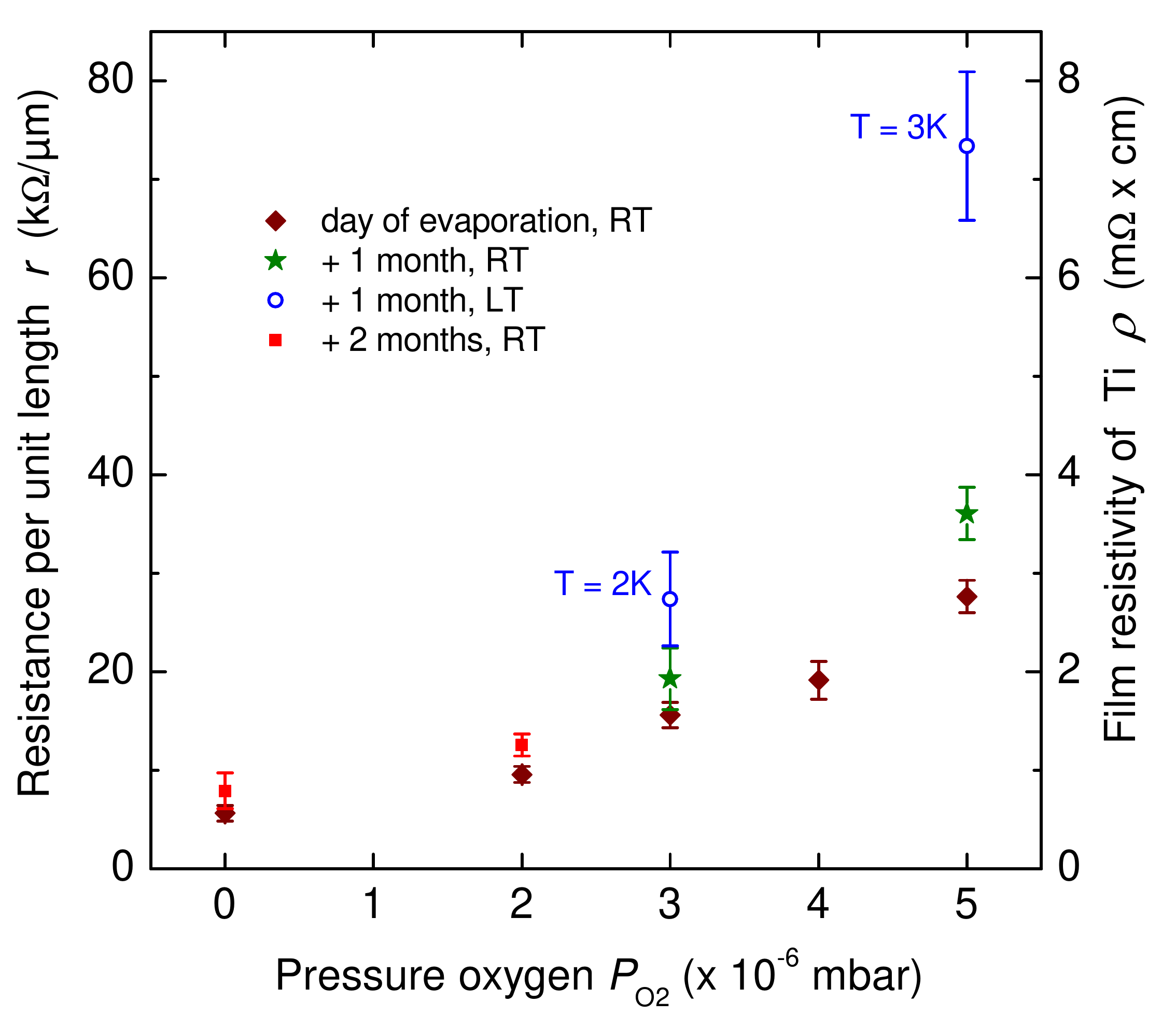}
\caption{The resistance of Ti film as a function of oxygen pressure during evaporation.
Brown diamonds and error bars, respectively, represent a sample average and a standard deviation over
several different microstrips measured on the day of evaporation.
For comparison, we also show the same data obtained after prolonged exposure of the samples to air, at room
(RT, red squares and green stars) and at low (LT) temperatures (blue empty circles).}
\label{O2}
\end{figure}

The experimental structures, see Fig.~\ref{SEM}, were fabricated using
electron-beam lithography and the shadow-evaporation technique
\cite{Dolan} which is often implemented for structuring ultrasmall tunnel
junctions of Al (see a detailed description of our fabrication routine in
Ref.~\cite{LotkhovNMDC}). In this technique, a $\unit[10]{nm}$ thin Ti
film was first evaporated from the standard e-gun source at a slanted
incidence of material onto the substrate, followed by a normal-angle
evaporation of a thicker film of AuPd to form well-conducting current
leads. Because of the well-pronounced gettering property of Ti films, the
important parameters of our process were the base chamber pressure, $p <
\unit[5 \times 10^{-9}]{mbar}$, and the evaporation rate of Ti
$\unit[0.2]{nm/s}$. In order to increase film resistivity, the deposition
of Ti was performed reactively with a small amount of oxygen gas added to
the evaporation chamber at a low pressure $P_{\rm {O2}} \le \unit[5 \times
10^{-6}]{mbar}$.  The samples were tested electrically at room temperature
on the day of the evaporation, demonstrating a clear monotonous dependence
of the resistance $r$ on the oxygen pressure, as shown in Fig.~\ref{O2}.
We selectively tested a few samples after a month or two of shelf life to
find an up to $\sim 50 \%$ increase in film resistivity due to the
prolonged air contact.

\begin{figure}[]
\centering%
\includegraphics[width=1\columnwidth]{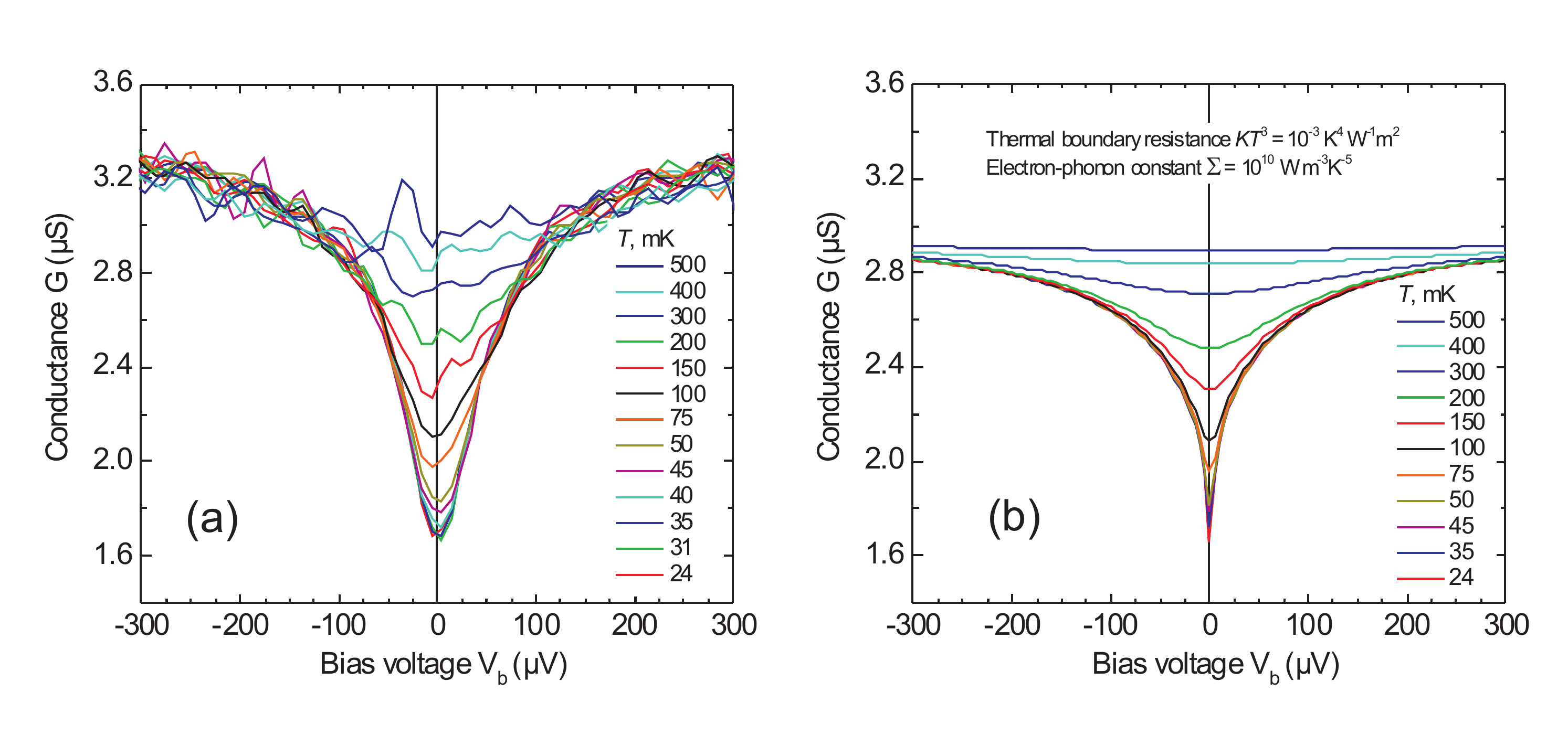}
\caption{Zero-bias conductance dip as a function of temperature: (a) measured experimentally and (b) calculated with the parameter values $K$ and $\Sigma$ shown in the plot.
We note that neither the value of $\Sigma$ nor that of $K$ was found to significantly influence the results of calculation. A somewhat higher estimated value of $\Sigma$
for high-resistivity Ti is found to be plausible, if compared with those for pure materials (cf., e.g., Refs.~\cite{Roukes1985,Wellstood1994}).}
\label{dip}
\end{figure}

The resistors were characterized in the dilution fridge down to the
operating temperatures of the Coulomb blockade devices below
$\unit[100]{mK}$. Due to the oxygen content in the Ti film, a negative
temperature profile of resistance was observed with the values increasing
by the factor 1.5---2 down to a few degrees Kelvin, see Fig.~\ref{O2}.
This temperature behavior is consistent with the findings of other groups
reported for high-ohmic Ti films, see, e.g.,
Refs.~\cite{Singh1972,Hoffmann2003}, indicating a strong impurity
scattering limit of conductivity electrons in the film. The $I-V_{\rm b}$
curves were found to be almost linear for all tested resistors, with a
weak nonlinearity around the origin, appearing at the lowest temperatures
as a zero-bias conductance dip $G(V_{\rm b},T)$ in Fig.~\ref{dip}(a).
Remarkably, the height of the dip, increasing towards lower temperatures,
was found to come to saturation at the substrate temperature $T \approx
\unit[35]{mK}$ very close to the base temperature of the fridge. This
indicates very efficient thermalization of the conduction electrons on the
phonons of the crystal lattice and gives rise to a promise of a very low
level ($\propto T$) of thermal noise in the mK-temperature range, which is
quite favorable for applications in the quantum current standard
\cite{LikharevZorin,MooijNazarov,LotkhovRpump}.

For a more detailed understanding of the current transport mechanism, we
first address a straightforward hypothesis of a purely thermal origin of
the conductance dip, $G(V_{\rm b},T) = G[T_{\rm e}(V_{\rm b},T),T_{\rm
{ph}}(V_{\rm b},T)]$. We take into account the Joule heating effect of
current through the Ti resistor and calculate the electron temperature
$T_{\rm e}$ according to the one-fifth power law
\cite{Roukes1985,Wellstood1994}:

\begin{equation}
T_{\rm e} = \left[T_{\rm {ph}}^5 + \frac{P}{\Sigma\Omega}\right]^{1/5},\label{eq:powerlaw}
\end{equation}
where $P$ is the heat flow from the electron gas to the lattice phonons in
the Ti microstrip, $\Sigma$ is the electron-phonon coupling constant,
$\Omega = \unit[0.01 \times 0.1 \times 4]{\mu m^3}$ the volume of the
resistor, and $T_{\rm {ph}} = T + KP/A$ the phonon (lattice) temperature
of the resistor. Here, $A = \unit[0.1 \times 4]{\mu m^2}$ is the contact
area between the microstrip and the substrate. We expect $T_{\rm {ph}}
\approx T$  to be close to the substrate temperature due to a typically
small value of thermal boundary resistance $K$, namely, $K \times T^3 <
\unit[0.001]{K^4W^{-1}m^2}$ \cite{Swartz1989}.

An important simplification of our model was made due to the high
resistivity of the Ti film giving rise to a strong scattering
approximation for conduction electrons on the crystal lattice. In
particular, we assumed a fully local heat dissipation by the
electron-phonon interaction mechanism and neglected small energy losses
across the boundary Ti/AuPd, so that $P \approx IV_{\rm b}$.
Correspondingly, the temperature profile along the microstrip is regarded
as being flat except for the sharp temperature drops at both ends of the
Ti wire. Further on, at $V_{\rm b} = 0$ we assumed $T_{\rm e} = T_{\rm
{ph}}= T$ and used the experimental zero-bias conductance values  to
establish an empirical calibration relation $G(T_{\rm e} = T) = G(0,T)$.
From now on, we adopt $G$, being independent of $T_{\rm {ph}}$: the
approximation is justified in part by an independence of $G$ on $T$ in
Fig.~\ref{dip}(a) for $|{V_{\rm b}}| > \unit[100]{\mu V}$, where $T_{\rm
e} \approx \left[\frac{P}{\Sigma\Omega}\right]^{1/5}$ is common for all
curves, but the values of $T_{\rm {ph}} \approx T$ vary. Finally, the
conductance dips $G(V_{\rm b},T)$ were calculated to be shown in
Fig.~\ref{dip}(b), using the calibration curve $G(T_{\rm e})$ and the
values $T_{\rm e} = T_{\rm e}(V_{\rm b},T)$ obtained from
Eq.~\ref{eq:powerlaw}.

The assumption $T_{\rm e} = T_{\rm {ph}}= T$ used for the calibration
curve $G(T_{\rm e})$ ignores possible additional heating of the sample by
the external noise (voltage fluctuations coming through the leads to the
Ti microstrip) and supposedly results in the sharper and less noisy dips
calculated for the lowest values of $T$ in Fig.~\ref{dip}(b). On the other
hand, our even simplified model demonstrates a reasonable agreement with
the experiment. In this way, it provides a distinct argument for the
purely thermal origin of the dip, making less probable a presence of
unwanted granular Coulomb blockade effects (cf., e.g., the case of
partially oxidized Cr films \cite{KrupeninChromium} or the anodized Ti
films in Ref.~\cite{Johannson2000}) and related shot-noise effects. An
indirect support of this conclusion also appears in the form of a very low
noise level measured in the resistor in the frequency range around $f =
\unit[10]{Hz}$, typical for the background charge fluctuations phenomena
in the single-electron tunneling devices \cite{Krupenin2000}. The noise
was found unresolvable against the intrinsic noise level $\delta I \sim
\unit[10^{-13}]{A/\sqrt{Hz}}$ of the current preamplifier.

Due to the higher attainable values of $r$ in the Ti, as compared to Cr,
and according to Eq.~\ref{eq:rolloff}, a roll-off frequency higher by one
order of magnitude is now expected for the high-ohmic environmental
impedance available for applications. For example, for the hybrid
single-electron turnstile \cite{Pekola-NaturePhysics2008} with typically
$R_0 \sim \unit[100]{k\Omega}$ \cite{Lotkhov2009}, the roll-off frequency
is $\omega_0 \sim \unit[10^{11}]{s^{-1}}$ and the high-impedance behavior
extends up to $\omega_{\rm c} \sim \unit[10^{12}]{s^{-1}}$ \cite{Ct},
indicating a remarkably wide energy range, up to $\hbar\omega \sim
\unit[1]{meV}$, where the quantum system is decoupled from the
environment, thus making the suppression of Cooper-pair-electron (CPE)
cotunneling leakage \cite{Pekola-NaturePhysics2008} potentially more
efficient. For applications with the Bloch oscillations devices
\cite{KuzminHaviland,Maibaum2012} and quantum phase slip junctions
\cite{HongistoZorin}, a narrowing by an order of magnitude for the
linewidth of oscillations, $\Gamma \propto T_{\rm e}/R_0$
\cite{LikharevZorin} and, therefore, substantial improvement of the shape
of Shapiro-like current steps at $I = 2ef$ (under microwave irradiation of
frequency $f$, in the limit $\Gamma / f \to 0$)  are expected
\cite{LikharevBlueBook}.

In conclusion, we report on exceedingly high-ohmic microstripline
resistors made of titanium films evaporated  reactively in the presence of
oxygen. This new material exhibited a sheet resistance of up to
$\unit[7]{k\Omega}$ which is an order of magnitude higher than that of
formerly used Cr films and is expected to be free of intrinsic Coulomb
blockade effects related to film granularity even at the lowest
temperatures of the experiment. Due to the latter fact, a further increase
in resistivity seems to be possible without immediate loss of linearity.
The obtained parameters, in particular, a considerable improvement in the
roll-off frequency, are promising for the application of Ti resistors in
different realizations of quantum standards of current.

\section*{Acknowledgements}

Stimulating discussions with A.~B.~Zorin are gratefully acknowledged.
Technological support from T.Weimann and V.~Rogalya is appreciated.


\section*{References}
\begin{thebibliography}{26}


\bibitem{CaldeiraLeggett} Caldeira A O and Leggett A J 1983 Quantum
    tunnelling in a dissipative system {\it Annals of Physics} {\bf 149}
    374-456

\bibitem{IngoldNazarov} Ingold G L and Nazarov Yu V 1992 Charge tunneling
    rates in ultrasmall junctions {\it Single Charge Tunneling, Coulomb Blockade Phenomena in Nanostructures} (NATO
    ASI Series B vol~294)(New York: Plenum Press) ch 2

\bibitem{LikharevZorin} Likharev K K and Zorin A B 1985 Theory of the
    Bloch-wave oscillations in small Josephson junctions {\it J Low Temp
    Phys } {\bf 59} 347-381

\bibitem{Pekola-NaturePhysics2008} Pekola J P, Vartiainen J J,
    M\"ott\"onen M, Saira O-P, Meschke M, and Averin D  V 2008 Hybrid single-electron transistor
    as a source of quantized electric current {\it Nature Physics} {\bf
    4} 120

\bibitem{MooijNazarov} Mooij J E and Nazarov Yu V Quantum phase slip
    junctions 2006 {\it Nature Phys.} {\bf 2} 169-172

\bibitem{KuzminHaviland} Kuzmin L S and Haviland D B 1991 Observation of
    the Bloch oscillations in an ultrasmall Josephson Junction {\it Phys Rev
    Lett} {\bf 67} 2890-2893

\bibitem{LotkhovRpump} Lotkhov S V, Bogoslovsky S A, Zorin A B and
    Niemeyer J 2001 Operation of a three-junction single-electron pump with on-chip resistors {\it Appl Phys Lett } {\bf 78}
    946-948

\bibitem{LotkhovBlochArray} Lotkhov S V, Krupenin V A, and Zorin A B 2007
    Cooper pair transport in a resistor-biased Josephson junction array
    {\it IEEE Trans Instr Meas} {\bf 56} 491-494

\bibitem{Maibaum2012} Maibaum F, Lotkhov S V, and Zorin A B 2011 Towards
    the observation of phase-locked Bloch oscillations in arrays of small
    Josephson junctions {\it Phys Rev B} {\bf 84} 174514

\bibitem{CamarotaMetrologia} Camarota B, Scherer H, Keller M W, Lotkhov S
    V, Willenberg G-D, and Ahlers F J 2012 Electron counting capacitance
    standard with an improved five-junction R-pump {\it Metrologia} {\bf
    49} 8-14

\bibitem{HongistoZorin} Hongisto T T and Zorin A B 2012 Single-charge
    transistor based on the charge-phase duality of a superconducting
    nanowire circuit {\it Phys Rev Lett} {\bf 108} 097001


\bibitem{ZorinJAP2000} Zorin A B, Lotkhov S V, Zangerle H, and Niemeyer J
    2000 Coulomb blockade and cotunneling in single electron circuits with
    on-chip resistors: Towards the implementation of the R-pump{\it J Appl Phys } {\bf 88}
    2665-2670

\bibitem{KrupeninChromium} Krupenin V A, Zorin A B, Savvateev M N, Presnov
    D E, and Niemeyer J 2001 Single-electron transistor with metallic microstrips
    instead of tunnel junctions {\it J. Appl. Phys.} {\bf 90} 2411

\bibitem{Webelements} see, e.g., online:
    http://www.webelements.com/titanium/physics.html

\bibitem{Dolan} Dolan G J 1977 Offset masks for lift-off
    photoprocessing {\it Appl Phys Lett} {\bf 31} 337-339,
    Niemeyer J 1974 Eine einfache Methode zur Herstellung kleinster
    Josephson-Elemente {\it PTB-Mitt} {\bf 84} 251-253

\bibitem{LotkhovNMDC} Lotkhov S V, Camarota B, Scherer H, Weimann T, Hinze
    P, and Zorin A B 2009 Shunt-protected single-electron tunneling circuits fabricated on a quartz
    wafer {\it Nanotechnology Materials and Devices Conference, NMDC '09 IEEE} 23 - 26

\bibitem{Singh1972} Singh R and  Surplice N A 1972 The electrical
    resistivity and resistance-temperature characteristics of thin
    titanium films {\it Thin  Solid  Films} {\bf 10} 243-253

\bibitem{Hoffmann2003} Hofmann K, Spangenberg B, Luysberg, and M Kurz H
    2003 Properties of evaporated titanium thin films and their possible
    application in single electron devices {\it Thin  Solid  Films} {\bf 436} 168–174

\bibitem{Roukes1985} Roukes M L, Freeman M R, Germain R S, and Richardson
    R C 1985 Hot electrons and energy transport in metals at millikelvin
    temperatures {\it Phys Rev Lett} {\bf 55} 422-425

\bibitem{Wellstood1994} Wellstood F, C Urbina C, and Clarke J 1994
    Hot-electron effects in metals {\it Phys Rev B} {\bf 49} 5942-5955

\bibitem{Swartz1989} Swartz E T and Pohl R O 1989 Thermal boundary
    resistance {\it Rev Mod Phys} {\bf 61} 605-668

\bibitem{Johannson2000} Johannson J, Sch\"ollmann V, Andersson K, and
    Haviland D B 2000 Coulomb blockade in anodised titanium nanostructures
    {\it Physica B} {\bf 284-288} 1796-1797

\bibitem{Krupenin2000} Krupenin V A, Presnov D E, Zorin A B, and Niemeyer
    J 2000 Aluminium single electron transistors with islands isolated from
    the substrate {\it J. Low Temp. Phys.} {\bf 118} 287-296

\bibitem{Lotkhov2009} Lotkhov S V, Kemppinen A, Kafanov S, Pekola J P and
    Zorin A B 2009 Pumping properties of the hybrid single-electron transistor
    in dissipative environment {\it Appl Phys Lett } {\bf 95} 112507

\bibitem{Ct} This value should however be reduced, if taking into account
    the tunnel junction capacitances contributing to the total effective impedance
    of the environment. Still, for the state-of-the-art junctions with typically $C_{\rm t}
    \sim \unit[50]{aF}$, this correction will not be critical.

\bibitem{LikharevBlueBook} Likharev K K 1986 Dynamics of Josephson
    Junctions and Circuits ({\it Gordon and Breach, New York})


\end{thebibliography}
\end{document}